THE EUROPEAN
PHYSICAL JOURNAL C

Regular Article - Theoretical Physics

# A simple ghost free and caustic free mimetic scalar field dark matter model

Emmanuel Kanambaye[a]

Librairie BAH SARL, Hall du Grand Hôte de Bamako-Mali, BP 104, Bamako, Mali

**Abstract** Chamseddine and Mukhanov have promoted the concept of mimetic dark matter as alternative dark matter candidate coming from gravitation. Unfortunately, although being very interesting, their proposition turned out to have weaknesses among which ghost and caustic instabilities. In this paper, I propose among the simplest ghost-free and caustic-free scalar field dark matter extension of their model; a proposition capable of challenging $\Lambda$CDM model or even capable of doing better.

## 1 Introduction

In the paper [1], Chamseddine and Mukhanov have proposed the concept of mimetic dark matter.

Their proposition is interesting in the sense that in their framework dark matter comes from gravitation as natural consequence of a conformal extension of general relativity [2] with local Weyl invariance in terms of the fundamental metric.

In [3], Golovnev showed that the model proposed by Chamseddine and Mukhanov is equivalent to the use of a Lagrange multiplier constraining the kinematic of some scalar field known as the mimetic field.

However although very interesting [4], the model [1,3] appeared to have ghost and caustic instabilities, see papers [5,6] for details.

In the present paper, I show by using a trick involving "constrained auxiliary fields", how a ghost free and caustic free scalar field dark matter extension of the model proposed in [1,3] can be achieved.

As we will see, the gotten model is among the simplest scalar field dark matter model [7,8], yielding stiff matter [9,

10] and "cosmological constant" dark energy; a model quite capable of reproducing $\Lambda$CDM model [11] or doing better.

The paper is organised as follow, in the Sect. 2, I make a brief recall of original mimetic gravity and its rather simple Friedmann cosmology as well as its weakness; In Sect. 3, I derive the new model from variational principle; In Sect. 4, I discuss its cosmology and show that it yields beside a stiff matter, a simple caustics free and ghost free dark matter candidate of which the density has some connection with dark energy density; In Sect. 5, I show how coupling of the dark matter sector with inflaton field can be achieved in our model, while I devote Sect. 6 for conclusion.

Also the signature convention $(+ - - -)$ will be assumed for the metric; likewise Einstein's summation convention as well as natural unit for the speed of light in vacuum i.e. $c = 1$ will be assumed.

## 2 Original mimetic gravity

In their paper [1], Chamseddine and Mukhanov proposed to isolate the conformal degree of freedom of general relativity [2] by introducing a parametrization of the physical metric $g_{\mu\nu}$ as follows:

$$g_{\mu\nu} = \left(\tilde{g}^{ab}\partial_a\phi\partial_b\phi\right)\tilde{g}_{\mu\nu} \qquad (1)$$

where $\tilde{g}_{\mu\nu}$ is an auxiliary metric and $\phi$, a scalar field that they called "mimetic field".

The interest of such definition (1), is that the physical metric $g_{\mu\nu}$ would be invariant with respect to the conformal transformation $\tilde{g}_{\mu\nu} \to \Omega^2 \tilde{g}_{\mu\nu}$ of the auxiliary metric.

Likewise, the scalar field $\phi$ should satisfy the constraint:

$$g^{\mu\nu}\partial_\mu\phi\partial_\nu\phi = 1. \qquad (2)$$

The gravitational action of the theory should therefore be constructed in terms of the physical metric, which will be

[a] e-mail: wadouba@gmail.com (corresponding author)

Springer

considered as a function of the scalar field and the auxiliary metric; which in the presence of standard matter of lagrangian $2\kappa L_\Psi$ would read:

$$S = -\frac{1}{2} \int \left( R(\tilde{g}_{\mu\nu}, \phi) + 2L_\Psi \right) \sqrt{-g(\tilde{g}_{\mu\nu}, \phi)} d^4x \quad (3)$$

where we have set the Einstein gravitational constant $\kappa = 1$; with $R(\tilde{g}_{\mu\nu}, \phi)$ the Ricci scalar and $g(\tilde{g}_{\mu\nu}, \phi)$ the determinant of the physical metric.

Now as discussed in [1], variation of (3) with respect to the physical metric by taking into account its dependence from the auxiliary metric and the scalar field $\phi$ would yield the gravitational field equation reading:

$$G_{\mu\nu} = T_{\mu\nu} + [G - T]\partial_\mu \phi \partial_\nu \phi \quad (4)$$

$$\nabla^\mu \left[ [G - T]\partial_\mu \phi \right] = 0 \quad (5)$$

where $G$ and $T$ are the trace of respectively $G_{\mu\nu}$ and $T_{\mu\nu}$; with $G_{\mu\nu}$ the Einstein tensor and $T_{\mu\nu} = -\frac{2}{\sqrt{-g}} \frac{\delta(\sqrt{-g}L_\Psi)}{\delta g^{\mu\nu}}$ the energy–momentum tensor of standard matter.

As we see, the auxiliary metric does not enter the gravitational field equation, at least not explicitly, while the mimetic field enters explicitly.

The Eqs. (2), (4) and (5) constitute thus what we know as mimetic gravity [4].

Let's note that soon after Chamseddine and Mukhanov work [1]; Golovnev showed [3] that the model has an equivalent formulation from the action:

$$S = \frac{1}{2} \int \left( -R + 2L_\Psi + \lambda \left[ \partial_\mu \phi \partial^\mu \phi - 1 \right] \right) \sqrt{-g} d^4x \quad (6)$$

where R is the Ricci Scalar; $L_\Psi$ the lagrangian of standard matter fields [12]; $\lambda$ a Lagrange multiplier and $\phi$ the mimetic field.

Indeed applying variational principle to this action (6) would yield the field equations:

$$G_{\mu\nu} = T_{\mu\nu} + \lambda \partial_\mu \phi \partial_\nu \phi \quad (7)$$

$$g^{\mu\nu} \partial_\mu \phi \partial_\nu \phi = 1 \quad (8)$$

$$\nabla^\mu \left[ \lambda \partial_\mu \phi \right] = 0. \quad (9)$$

Now as the trace of the Eq. (7) yields $\lambda = G - T$, it results that we reproduce exactly the mimetic gravity [1] of Chamseddine and Mukhanov.

It is obvious that $\lambda \partial_\mu \phi \partial_\nu \phi$ can be interpreted as the energy–momentum tensor of a perfect pressureless fluid of energy-density $\lambda$ and four-velocity $\partial_\mu \phi$.

Let's note that the principal interest of mimetic gravity (3)/(6), is that it naturally yields simple dark matter candidate.

Indeed in cosmology, more precisely in the case of spatially homogeneous and isotropic Friedmann metric [13] of line-element:

$$ds^2 = dt^2 - \gamma_{ij} dx^i dx^j \quad (10)$$

where $\gamma_{ij}$ denotes the three dimensional spatial metric; the Eq. (4)/(7) would by assuming perfect fluid standard matter, yield the following Friedmann equations:

$$3H^2 + 3\frac{K}{a^2} = \rho_\Psi + \lambda \dot{\phi}^2 \quad (11)$$

$$2\dot{H} + 3H^2 + 3\frac{K}{a^2} = -P_\Psi \quad (12)$$

where $H = \frac{\dot{a}}{a}$ denotes the Hubble parameter; $a$ the scale factor; $K/a^2$ the spatial curvature; $\rho_\Psi$ and $P_\Psi$ respectively the energy-density and pressure of standard matter; $\dot{\phi} = 1$ and $\lambda = \frac{C}{a^3}$ with $C$ a constant of integration.

Thus, it appears that $\lambda$ contributes as pressureless dust dubbed mimetic dark matter [1], which is among the simplest way of achieving dark matter from gravitation.

Unfortunately, although being very interesting; Chamseddine and Mukhanov mimetic gravity appears having weakness among which caustics and ghost instabilities as very well exposed in the papers [5,6].

Clearly, the formation of caustics is due to the fact that mimetic dark matter is a perfect pressureless dust moving along geodesics [5], while the ghost instability comes from the possibility for $\lambda$ to evolve from positive value towards negative one, more precisely towards negative infinite [5,6] and thus making unbounded from below the Hamiltonian of the mimetic field $\phi$.

Therefore to cure mimetic gravity; we should succeed in constraining $\lambda$ to not yield Hamiltonian unbounded from below, while endowing the dark matter with for example some pressure capable of precluding caustics formation by making the dark matter particles not exactly moving along geodesics [14–16].

This is what we are going to do in the next section; more precisely, by a clever use of other Lagranges multipliers and auxiliary/mathematical scalar field $\chi$, I succeed (by involving additional canonical scalar field $\psi$) in constraining $\lambda$ to ensure Hamiltonian bounded from below, beside the mimetic condition (8), while endowing the obtained dark matter with some pressure capable of precluding caustics formation [14–16].

Let's see how.

## 3 The model

To get a ghost free and caustic free scalar field dark matter version/extension of the model proposed in [1,3], we can consider the below modification/extension (13) of the action (6):

$$S = \frac{1}{2} \int \left( -R + 2L_\Psi + \lambda \left[ \partial_\mu \phi \partial^\mu \phi - 1 - \frac{\alpha}{\mathcal{A}} \right] \right.$$

$$+\varepsilon[\lambda - \mathcal{A}] + L_{\chi,\psi}\bigg)\sqrt{-g}d^4x \quad (13)$$

where we have set the Einstein gravitational constant $\kappa = 1$; with $R$ the Ricci scalar; $L_\Psi$ the lagrangian of standard matter fields [12]; $\lambda$ and $\varepsilon$ Lagrange multipliers; $\phi$ the mimetic field; $\mathcal{A} = \mathcal{A}(\partial\phi, \partial\psi)$ some function of $\partial\phi$ and $\partial\psi$, which will be determined later; with $\psi$ another canonical scalar field.

Of course as we will see later, the characteristics of the gotten dark matter would depend from the exact expression of the function $\mathcal{A} = \mathcal{A}(\partial\phi, \partial\psi)$.

Finally, we recall that $L_{\chi,\psi}$ denotes a lagrangian reading:

$$L_{\chi,\psi} = \alpha\partial_\mu\psi\partial^\mu\psi + \beta\partial_\mu\phi\partial^\mu\psi + \omega\big[\partial_\mu\chi\partial^\mu\chi - 1\big]$$
$$+\omega_k[\omega - \alpha] - \alpha\partial_\mu\chi\partial^\mu\chi + \alpha^{\frac{3}{2}}\chi\big[\partial_\mu\phi\partial^\mu\phi - 1\big]^2 \quad (14)$$

where $\alpha > 0$ and $\beta \neq 0$ are free constants; $\chi$ an auxiliary scalar field, while $\omega$ and $\omega_k$ are Lagrange multipliers.

Let's note that a potential term $V(\phi)$ for the mimetic field $\phi$, capable of ensuring amongst other things dynamical dark energy [4,17,18], can be added in (13); though this would require some subtlety; whence for simplicity, we will in the present work assume $V(\phi) = 0$.

Now as all the elements of our action (13) are specified; then we can apply variational principle.

- Indeed varying (13) with respect to $\delta g^{\mu\nu}$ would yield

$$G_{\mu\nu} = T_{\mu\nu} + \lambda\partial_\mu\phi\partial_\nu\phi + T^\psi_{\mu\nu} + \psi^\phi_{\mu\nu} - \lambda_{\mu\nu}$$
$$+[\omega - \alpha]\chi_{\mu\nu} + \frac{\omega}{2}g_{\mu\nu} + \alpha_{\mu\nu} \quad (15)$$

where $T_{\mu\nu} = -\frac{2}{\sqrt{-g}}\frac{\delta(\sqrt{-g}L_\Psi)}{\delta g^{\mu\nu}}$ denotes the energy–momentum tensor of standard matter fields; while:

$$\begin{cases} T^\psi_{\mu\nu} = \alpha\big[\partial_\mu\psi\partial_\nu\psi - \frac{1}{2}g_{\mu\nu}\partial_\sigma\psi\partial^\sigma\psi\big] \\ \psi^\phi_{\mu\nu} = \beta\big[\partial_\mu\phi\partial_\nu\psi - \frac{1}{2}g_{\mu\nu}\partial_\sigma\phi\partial^\sigma\psi\big] \\ \lambda_{\mu\nu} = \lambda\frac{\delta}{\delta g^{\mu\nu}}[\frac{\alpha}{\mathcal{A}}] + \varepsilon\frac{\delta\mathcal{A}}{\delta g^{\mu\nu}} \\ \quad +\frac{1}{2}g_{\mu\nu}\big[\lambda\partial_a\phi\partial^a\phi - \lambda - \lambda\frac{\alpha}{\mathcal{A}} + \varepsilon[\lambda - \mathcal{A}]\big] \\ \chi_{\mu\nu} = \partial_\mu\chi\partial_\nu\chi - \frac{1}{2}g_{\mu\nu}\partial_a\chi\partial^a\chi \\ \alpha_{\mu\nu} = \alpha^{\frac{3}{2}}\chi\big[\partial_\mu\phi\partial^\mu\phi - 1\big]\partial_\mu\phi\partial_\nu\phi \\ \quad -\frac{1}{2}\alpha^{\frac{3}{2}}\chi g_{\mu\nu}\big[\partial_a\phi\partial^a\phi - 1\big]^2 - \frac{1}{2}g_{\mu\nu}\omega_k[\omega - \alpha]. \end{cases}$$

- Likewise, varying (13) with respect to $\delta\lambda$, $\delta\varepsilon$, $\delta\psi$ and $\delta\phi$, would respectively yield:

$$\partial_\mu\phi\partial^\mu\phi - 1 - \frac{\alpha}{\mathcal{A}} + \varepsilon = 0 \quad (16)$$

$$\lambda = \mathcal{A} \quad (17)$$

$$\nabla^\mu\bigg[2\partial_\mu\psi + \beta\partial_\mu\phi + \lambda\frac{\alpha}{\mathcal{A}^2}\frac{\partial\mathcal{A}}{\partial(\partial^\mu\psi)} - \varepsilon\frac{\partial\mathcal{A}}{\partial(\partial^\mu\psi)}\bigg] = 0 \quad (18)$$

$$\nabla^\mu\bigg[2\lambda\partial_\mu\phi + \beta\partial_\mu\psi + \lambda\frac{\alpha}{\mathcal{A}^2}\frac{\partial\mathcal{A}}{\partial(\partial^\mu\phi)} - \varepsilon\frac{\partial\mathcal{A}}{\partial(\partial^\mu\phi)}$$
$$+4\alpha^{\frac{3}{2}}\chi\big[\partial_\mu\phi\partial^\mu\phi - 1\big]\partial_\mu\phi\bigg] = 0 \quad (19)$$

- while varying (13) with respect to $\delta\omega$, $\delta\omega_k$ and $\delta\chi$ would respectively yield:

$$\partial_\mu\chi\partial^\mu\chi - 1 + \omega_k = 0 \quad (20)$$
$$\omega - \alpha = 0 \quad (21)$$
$$\nabla^\mu\big[2\omega\partial_\mu\chi - 2\alpha\partial_\mu\chi\big] - \alpha^{\frac{3}{2}}\big[\partial_\mu\phi\partial^\mu\phi - 1\big]^2 = 0. \quad (22)$$

From there, it is clear that because of (21); the Eq. (22) would simplify into:

$$\alpha^{\frac{3}{2}}\big[\partial_\mu\phi\partial^\mu\phi - 1\big]^2 = 0 \quad (23)$$

what enforces:

$$\partial_\mu\phi\partial^\mu\phi - 1 = 0. \quad (24)$$

This latter (24) would in its turn because of (16), yield for the lagrange multiplier $\varepsilon$, the solution:

$$\varepsilon = \frac{\alpha}{\mathcal{A}}. \quad (25)$$

In the same way, this above (25) would in its turn because of (17), imply:

$$\lambda\frac{\alpha}{\mathcal{A}^2}\frac{\partial\mathcal{A}}{\partial(\partial^\mu\psi)} - \varepsilon\frac{\partial\mathcal{A}}{\partial(\partial^\mu\psi)} = 0 \quad (26)$$

$$\lambda\frac{\alpha}{\mathcal{A}^2}\frac{\partial\mathcal{A}}{\partial(\partial^\mu\phi)} - \varepsilon\frac{\partial\mathcal{A}}{\partial(\partial^\mu\phi)} = 0 \quad (27)$$

$$\lambda_{\mu\nu} = -\frac{1}{2}\alpha g_{\mu\nu} \quad (28)$$

$$\alpha_{\mu\nu} = 0 \quad (29)$$

whence in summary, by considering (21) as well as (24) which enforces $\nabla^\mu\big[\beta\partial_\mu\phi\big] = 0$; the Eqs. (15), (16), (17), (18) and (19) could respectively simplify as:

$$G_{\mu\nu} = T_{\mu\nu} + \lambda\partial_\mu\phi\partial_\nu\phi + \psi^\phi_{\mu\nu} + T^\psi_{\mu\nu} + \alpha g_{\mu\nu} \quad (30)$$
$$\lambda = \mathcal{A} \quad (31)$$
$$\partial_\mu\phi\partial^\mu\phi - 1 = 0 \quad (32)$$
$$\nabla^\mu\nabla_\mu\psi = 0 \quad (33)$$
$$\nabla^\mu\big[\lambda\partial_\mu\phi\big] = \nabla^\mu\big[\mathcal{A}\partial_\mu\phi\big] = 0 \quad (34)$$

where of course $T_{\mu\nu} = -\frac{2}{\sqrt{-g}}\frac{\delta(\sqrt{-g}L_\Psi)}{\delta g^{\mu\nu}}$ denotes the energy–momentum tensor of standard matter; $\psi^\phi_{\mu\nu} = \beta\big[\partial_\mu\phi\partial_\nu\psi - \frac{1}{2}g_{\mu\nu}\partial_\sigma\phi\partial^\sigma\psi\big]$ and $T^\psi_{\mu\nu} = \alpha\big[\partial_\mu\psi\partial_\nu\psi - \frac{1}{2}g_{\mu\nu}\partial_\sigma\psi\partial^\sigma\psi\big]$.

Let's also notice that for a given standard matter field $\Psi$; its Euler–Lagrange equation of motion would be:

$$\nabla^\mu \left[ \frac{\partial L_\Psi}{\partial(\partial^\mu \Psi)} \right] - \frac{\partial L_\Psi}{\partial \Psi} = 0. \tag{35}$$

As we clearly see, the field $\chi$ as well as the lagrange multiplier $\omega_k$ contribute neither in the gravitation equation (30), nor in the equations of motion of the fields $\phi$, $\psi$ and $\Psi$; whence they can be seen as just auxiliary/mathematical fields (i.e. non-physical fields) helping to achieve the Euler–Lagrange equations of motion (30), (32), (33) and (34). In the same way, the Lagrange multiplier $\varepsilon$ also is excluded from the equations of motion of the theory thanks to the constraints (24) and (25), while $\lambda$ participates through $\mathcal{A}$, see (17)/(31); in the same way, we could say that $\omega$ also participates through $\alpha$.

Be that as it may, what is interesting with our gravitational theory (30–34) in comparison to original mimetic gravity (7–9); is that it is not only a constrained "system/theory" derivable from variational principle but also, it is (as we will see in the next section), among the simplest ghost free and caustic free "mimetic" gravity theory capable of challenging $\Lambda$CDM model [11] and/or doing better by yielding scalar field dark matter candidate [7,8], plus stiff matter [9,10], plus dark energy through $\alpha g_{\mu\nu}$.

Let's note in passing that the equation of motion (32), (33) and (34) ensures the below conservation equation (36):

$$\nabla^\mu \left[ \lambda \partial_\mu \phi \partial_\nu \phi + \psi^\phi_{\mu\nu} + T^\psi_{\mu\nu} \right] = 0 \tag{36}$$

whence the Bianchi identity $\nabla^\mu G_{\mu\nu} = 0$ would ensure the standard $\nabla^\mu T_{\mu\nu} = 0$.

## 4 Ghost free and caustic free mimetic scalar field dark matter

To study the cosmology of our model; let's consider the Friedmann metric line-element (10) characterizing spatially homogeneous and isotropic spacetime.

Likewise let's assume that our standard matter energy–momentum tensor $T_{\mu\nu}$ is the one of perfect fluid, reading:

$$T_{\mu\nu} = [\rho_\Psi + P_\Psi] u_\mu u_\nu - P_\Psi g_{\mu\nu} \tag{37}$$

where $\rho_\Psi$, $P_\Psi$ and $u_\mu$ are respectively the energy-density, the pressure and four-velocity of standard matter.

In that case where $\partial_j \phi = \partial_j \psi = 0$ etc. because of the spatial homogeneity and isotropy; the Eq. (30) would yield the following Friedmann equations:

$$3H^2 + 3\frac{K}{a^2} = \rho_\Psi + \lambda \dot{\phi}^2 + \frac{1}{2}\beta \dot{\phi}\dot{\psi} + \frac{1}{2}\alpha \dot{\psi}^2 + \alpha \tag{38}$$

$$2\dot{H} + 3H^2 + 3\frac{K}{a^2} = -P_\Psi - \frac{1}{2}\beta \dot{\phi}\dot{\psi} - \frac{1}{2}\alpha \dot{\psi}^2 + \alpha \tag{39}$$

where $H = \frac{\dot{a}}{a}$ denotes the Hubble parameter; $a$ the scale factor; $\frac{K}{a^2}$ the spatial curvature, while $\dot{\phi}$ is the time derivative of $\phi$ and $\dot{\psi}$ the time derivative of $\psi$.

In the same way, one can verify that the Eqs. (32) and (33) would respectively yield:

$$\dot{\phi} = 1 \tag{40}$$

$$\dot{\psi} = \frac{\psi_0}{a^3} \tag{41}$$

where $\psi_0 > 0$ is a constant of integration; whence (38) and (39) would, by considering (31), simplify/rewrite as:

$$3H^2 + 3\frac{K}{a^2} = \rho_\Psi + \mathcal{A} + \frac{1}{2}\beta \frac{\psi_0}{a^3} + \alpha \frac{\psi_0^2}{a^6} + \alpha \tag{42}$$

$$2\dot{H} + 3H^2 + 3\frac{K}{a^2} = -P_\Psi - \frac{1}{2}\beta \frac{\psi_0}{a^3} - \alpha \frac{\psi_0^2}{a^6} + \alpha. \tag{43}$$

As we clearly see; the pressure of our dark matter fluid reads $P_{dm} = \frac{1}{2}\beta \frac{\psi_0}{a^3}$, while its total energy-density could be dependent from $\mathcal{A}$ since this last could for example involve additional dark matter component and/or counter-term for $\frac{1}{2}\beta \frac{\psi_0}{a^3}$.

As consequence the choice of appropriate expression for $\mathcal{A}$ is crucial to get experimentally competitive dark matter candidate.

The question is therefore whether it exists natural and/or simple expressions for $\mathcal{A}$ guaranteeing not only simple and very competitive dark matter candidate, but also guaranteeing positive Hamiltonians for $\phi$ and $\psi$.

Interestingly the response to this question is as we are going to see positive.

To be convinced, let's first calculate the Hamiltonians of $\phi$ and $\psi$ from (13).

Indeed as from the action (13); we can verify that the conjugate momentum of $\phi$ and $\psi$ would by considering (17), (24) and (25) respectively simplify as:

$$p_\phi = 2\lambda \partial_t \phi + \beta \partial_t \psi = 2\mathcal{A} \partial_t \phi + \beta \partial_t \psi \tag{44}$$

$$p_\psi = 2\alpha \partial_t \psi + \beta \partial_t \phi \tag{45}$$

then we will for $\phi$ and $\psi$ respectively get the Hamiltonians $H^\phi$ and $H^\psi$ reading:

$$H^\phi = p_\phi \partial^t \phi - \beta \partial_\mu \phi \partial^\mu \psi = 2\mathcal{A} \partial_t \phi \partial^t \phi - \beta \partial_j \psi \partial^j \phi \tag{46}$$

$$H^\psi = p_\psi \partial^t \psi - \alpha \partial_\mu \psi \partial^\mu \psi - \beta \partial_\mu \phi \partial^\mu \psi = \partial_t \psi \partial^t \psi \\ - \beta \partial_j \psi \partial^j \psi - \beta \partial_j \psi \partial^j \phi \tag{47}$$

which could rewrite as:

$$H^\phi = 2\mathcal{A} \dot{\phi}^2 + \beta \nabla \phi \nabla \psi \tag{48}$$

$$H^\psi = \alpha \dot{\psi}^2 + \alpha \left[ \nabla \psi \right]^2 + \beta \nabla \phi \nabla \psi \tag{49}$$

where $\nabla$ denotes spatial derivation (gradient).

Now since because of (24) we have $\dot{\phi}^2 \geq 1$ and $\nabla \phi \geq 0$; then it results that the Hamiltonian $H^\psi$ is bounded from below ($H^\psi \geq 0$); while $H^\phi$ will be bounded from below if for example $\mathcal{A}$ satisfies:

$$\mathcal{A} \geq -\frac{1}{2\dot{\phi}^2}\beta \nabla \phi \nabla \psi \qquad (50)$$

This means that so that our theory (30–34) be ghost free; we should choose for $\mathcal{A}$ an expression permitting to satisfy the above (50).

On the other hand, since experimental data [19] favors pressureless dark matter; then it is clear from (42) and (43) that so that our dark matter candidate be experimentally competitive; $\mathcal{A}$ should yields additional dark matter component largely dominating over $\frac{1}{2}\beta\frac{\psi_0}{a^3}$ so that the pressure of the dark matter appears largely negligible in comparison to its density, and of course, the only simple way to achieve this in FLRW cosmology, is that $\mathcal{A}$ satisfies:

$$\mathcal{A} \propto \frac{\psi_0}{a^3}. \qquad (51)$$

In other words, so that our theory (30–34) be a ghost free "mimetic" gravity yielding a caustic free and experimentally competitive dark matter candidate; the function $\mathcal{A}$ should satisfy (50) and (51).

Fortunately there are plethora of possible expressions for $\mathcal{A}$ permitting to satisfy (50) and (51); though one of the simplest is:

$$\mathcal{A} = \alpha \frac{\partial_\nu \psi \partial^\nu \phi}{\dot{\phi}^2} \tanh\left(\frac{1}{[\nabla\phi\nabla\psi]^2 \partial^\nu \phi \partial_\nu \psi}\right) - \frac{1}{2\dot{\phi}^2}\beta \nabla \phi \nabla \psi. \qquad (52)$$

Indeed in the above case (52); both (50) and (51) will be satisfied.

Especially, (48) will simplify into:

$$H^\phi = 2\alpha[\partial_\nu \psi \partial^\nu \phi] \tanh\left(\frac{1}{[\nabla\phi\nabla\psi]^2 \partial^\nu \phi \partial_\nu \psi}\right) \geq 0 \qquad (53)$$

which is clearly a Hamiltonian bounded from below; whence in the case of (52); our theory (30–34) will be ghost free.

Of course, to rigorously prove that the theory is "well and truly" ghost free; we have to perform a detailed Hamiltonian analysis of the theory, what we leave for future work.

In the same way, with (52); the Friedmann equations (42) and (43) would become:

$$3H^2 + 3\frac{K}{a^2} = \rho_\Psi + \rho_{dm} + \alpha\frac{\psi_0^2}{a^6} + \alpha \qquad (54)$$

$$2\dot{H} + 3H^2 + 3\frac{K}{a^2} = -P_\Psi - P_{dm} - \alpha\frac{\psi_0^2}{a^6} + \alpha \qquad (55)$$

where $\rho_{dm} = [\alpha + \frac{1}{2}\beta]\frac{\psi_0}{a^3}$ denotes dark matter energy-density; $P_{dm} = \frac{1}{2}\beta\frac{\psi_0}{a^3}$ dark matter pressure; $\alpha\frac{\psi_0^2}{a^6}$ stiff matter contribution and $\alpha$ dark energy density.

From there, it is easy to notice the simplicity of the model; for example as our dark matter pressure $P_{dm}$ depends from the free constant $\beta$; it results that for $\beta \ll \alpha$; we would have $P_{dm} \ll \rho_{dm}$ i.e. our dark matter would mimick cold dark matter, what means that we could rather easily reproduce $\Lambda$CDM model [11] since the stiff density $\alpha\frac{\psi_0^2}{a^6}$ should be important before radiation domination and become negligible later [9,10]; whence so that our model (54) and (55) be in accordance with observations [19]; we just have to choose appropriate values for the constants $\alpha$, $\beta$ and $\psi_0$.

Moreover, as our dark matter candidate is a scalar field dark matter [7,8] since coming from the coupling between the scalars fields $\psi$ and $\phi$, then it becomes quite possible to handle problems like small satellite problem [20] and/or cups-core problem [21] by assuming sufficiently light dark matter particle (see [7,8]).

Indeed, by posing $\rho_{dm} = [\alpha + \frac{1}{2}\beta]\frac{\psi_0}{a^3} = \frac{Nm_\psi}{a^3}$, where $N$ denotes the dark matter particles number and $m_\psi$ its mass; we would get $m_\psi = [\alpha + \frac{1}{2}\beta]\frac{\psi_0}{N}$; which, because of (33), gives us the freedom of choosing for the mass $m_\psi$ of our dark matter particle the value that suits us, and this, without great constraint from particle physics (in contrary to what happen in usual scalar field dark matter theories [7,8]) since in our paradigm, it is simply $[\alpha + \frac{1}{2}\beta]\dot{\phi}\dot{\psi}$ which yields dark matter density through (40) and (41) which latter put no constraints on what should be the mass of the dark matter particle.

In other words in our framework (30–34); we have a total freedom to assume for the mass of the dark matter particle the value that suits us in contrary to what happen in usual scalar field dark matter theories [7,8]; what makes our model (*in the case of sufficiently small mass for the dark matter particle*) quite capable of doing better than $\Lambda$CDM by explaining some weakness of this latter like small satellite problem [20] and/or cups-core problem [21].

It is also worth mentioning that since (in principle) our dark matter fluid has pressure ($\beta \neq 0$); then our dark matter particles would not exactly move along geodesics [14] and consequently would not form caustics singularities since the formation of caustics by dark matter is due to the fact that dark matter is assumed to be a perfect pressureless dust moving along geodesics [5]; in other words, in our paradigm (30–34), the pressure of the dark matter fluid would prevent the dark matter particles to cross with each other towards caustics singularities (see [15,16]); whence our dark matter model is caustic free!

We also see that, in our model; the dark matter density $\rho_{dm} = [\alpha + \frac{1}{2}\beta]\frac{\psi_0}{a^3}$ and the dark energy density are slightly connected to each other through the constant $\alpha$ which sets dark energy-density.

Because of this connection, our model seems more suitable than $\Lambda$CDM model, to explain the "cosmic coincidence problem" of why the current observed values of dark mat-ter density and dark energy density are of the same order of magnitude [22,23], since even though our dark matter density depends also from initial condition through the constant $\psi_0$; clearly dark matter density and dark energy-density appears more connected to each other in our paradigm than in $\Lambda$CDM model. Before ending this section; let's constraint the free parameters $\alpha$ and $\beta$ of our model on the basis of experimental data [19].

Indeed if we want that $\alpha$ plays the role of dark energy or cosmological constant; then experimental data [19] requires $\alpha \approx 1.088 \times 10^{-52} m^{-2}$.

In the same way, by posing $P_{dm} = w_{dm}\rho_{dm}$ with $w_{dm}$ the equation of state parameter of our dark matter fluid; we will get $w_{dm} = \frac{1}{2}\frac{\beta}{\alpha+\frac{1}{2}\beta}$.

Now as several experimental data studies [24–28] constrain dark matter equation of state parameter to satisfy in cosmology $w_{dm} < 10^{-4}$; then we can infer that our parameter $\beta$ should satisfy $\beta < \frac{2\alpha \times 10^{-4}}{1-10^{-4}} \approx 2.176 \times 10^{-56} m^{-2}$, what sets constraint on our non-zero dark matter pressure.

We thus see the simplicity of our dark matter model which could fit observations [19,24–28] for $\alpha \approx 1.088 \times 10^{-52} m^{-2}$ and $\beta < 2.176 \times 10^{-56} m^{-2}$.

It is clear at this stage that our model (30–34) has several interesting advantages in comparison to the original mimetic gravity (7–9), and we are going to see in the next section that these advantages remains practically unaffected in the case where we require interaction between the dark matter sector of the model and other field like for example inflaton field $\varphi$ [29,30].

## 5 Coupling with inflaton field

To couple our dark matter model to for example inflaton $\varphi$ [29,30]; we can add in the action (13) of our theory, the Lagrangian $L_\varphi$ reading:

$$L_\varphi = \partial_\mu \varphi \partial^\mu \varphi + V(\varphi) + [\phi + \psi]f(\varphi) \tag{56}$$

where $V(\varphi)$ is the potential of the inflaton field $\varphi$, and $f(\varphi)$ an appropriate function coupling the inflaton field to the fields $\phi$ and $\psi$.

Indeed in the case of such addition; we would instead of (30–34), have the theory:

$$G_{\mu\nu} = T_{\mu\nu} + \lambda \partial_\mu \phi \partial_\nu \phi + \psi^\phi_{\mu\nu} + T^\psi_{\mu\nu} + T^\varphi_{\mu\nu} + \alpha g_{\mu\nu} \tag{57}$$

$$\lambda = \mathcal{A} \tag{58}$$

$$\partial_\mu \phi \partial^\mu \phi - 1 = 0 \tag{59}$$

$$\alpha \nabla^\mu \nabla_\mu \psi - \frac{1}{2}f(\varphi) = 0 \tag{60}$$

$$\nabla^\mu \left[ \lambda \partial_\mu \phi + \frac{1}{2}\beta \partial_\mu \psi \right] - \frac{1}{2}f(\varphi) = 0 \tag{61}$$

$$\nabla^\mu \nabla_\mu \varphi - \frac{1}{2}\frac{dV(\varphi)}{d\varphi} - \frac{1}{2}[\phi + \psi]\frac{df(\varphi)}{d\varphi} = 0 \tag{62}$$

where $T^\varphi_{\mu\nu} = \left[\partial_\mu \varphi \partial_\nu \varphi - \frac{1}{2}g_{\mu\nu}\partial_\sigma \varphi \partial^\sigma \varphi\right] - \frac{1}{2}V(\varphi)g_{\mu\nu} - \frac{1}{2}g_{\mu\nu}[\phi + \psi]f(\varphi)$ is the energy–momentum tensor associated to the Lagrangian $L_\varphi$ given by (56).

Of course in the above case (57–62); instead of exactly considering for $\mathcal{A}$ the expression given by (52); it would be better to replace in (52) the constant $\alpha$ by the constant $\alpha - \frac{1}{2}\beta$ i.e. to consider:

$$\mathcal{A} = [\alpha - \frac{1}{2}\beta]\frac{\partial_\nu \psi \partial^\nu \phi}{\dot\phi^2} \tanh\left(\frac{1}{[\nabla\phi\nabla\psi]^2 \partial^\nu \phi \partial_\nu \psi}\right)$$
$$- \frac{1}{2\dot\phi^2}\beta \nabla\phi\nabla\psi \tag{63}$$

this replacement is needed otherwise we would get inconsistency between (60) and (61) in situations where $\nabla\phi\nabla\psi = 0$ like in FLRW cosmology.

Indeed in the case of (57–63), we would instead of the Friedmann equations (38) and (39), get:

$$3H^2 + 3\frac{K}{a^2} = \rho_\Psi + \alpha\dot\phi\dot\psi + \frac{1}{2}\alpha\dot\psi^2 + \rho_\varphi + \alpha \tag{64}$$

$$2\dot H + 3H^2 + 3\frac{K}{a^2} = -P_\Psi - \frac{1}{2}\beta\dot\phi\dot\psi - \frac{1}{2}\alpha\dot\psi^2 - P_\varphi + \alpha \tag{65}$$

where $\rho_\varphi$ and $P_\varphi$ are respectively the energy-density and pressure associated to $T^\varphi_{\mu\nu}$ and where $\dot\phi$ and $\dot\psi$ satisfy:

$$\dot\phi = 1 \tag{66}$$

$$\dot\psi = \frac{\psi_0}{a^3} + \frac{1}{2\alpha}\frac{1}{a^3}\int \left[a^3 f(\varphi)\right]dt \tag{67}$$

with $\psi_0 > 0$, a constant of integration.

From there, it is clear that for $\beta \ll \alpha$, we will retrieve (54) and (55) in regime where $f(\varphi)$ and/or $\varphi$ decays towards zero.

We can also mention that the coupling with inflaton $\varphi$, would for example not spoil the ghost free statute of the theory, as long as we assume appropriate coupling function $f(\varphi)$, appropriate potential $V(\varphi)$ and $\beta < 2\alpha$.

This comes from the fact that in the case of (57–63); we would instead of (53) and (49), have the Hamiltonians:

$$H^\phi = 2[\alpha - \frac{1}{2}\beta][\partial_\nu \psi \partial^\nu \phi] \tanh$$
$$\left(\frac{1}{[\nabla\phi\nabla\psi]^2 \partial^\nu \phi \partial_\nu \psi}\right) - \phi f(\varphi) \tag{68}$$

$$H^\psi = \alpha\dot\psi^2 + \alpha\left[\nabla\psi\right]^2 + \beta\nabla\phi\nabla\psi - \psi f(\varphi) \tag{69}$$

$$H^\varphi = \dot\varphi^2 + \left[\nabla\varphi\right]^2 - V(\phi) - [\phi + \psi]f(\varphi) \tag{70}$$

where $H^\varphi$ denotes the Hamiltonian of the inflaton field. Whence (57–63) would be ghost free for any appropriate $f(\varphi)$ and $V(\varphi)$.

## 6 Conclusion

We proposed in this paper among the simplest ghost free and caustic free "mimetic" scalar field dark matter model delivering also, stiff matter and "cosmological constant" dark energy; a model quite capable of challenging ΛCDM model if not doing better.

Indeed as we have seen, our dark matter model is a scalar field dark matter model in which we have a great freedom to choose for the mass of the dark matter particle the value that suits us more; as consequence it could (like usual scalar field dark matter model) rather easily handle some weakness of ΛCDM model like small satellite problem and/or cups-core problem as long as we assume a sufficiently light scalar dark matter particle.

It is also worth mentioning that the model seems more suitable than ΛCDM model to handle the cosmic coincidence problem since in our paradigm, the parameter $\alpha$ which sets the dark energy-density appears also in the dark matter density, what establishes some connection between dark matter density and dark energy-density.

Of course even though our dark matter density depends also from initial condition through the constant $\psi_0$; clearly dark matter density and dark energy-density appears more connected to each other in our paradigm than in ΛCDM model.

In view of all this; we can clearly say that our proposition (30–34), is a quite interesting dark matter model capable of challenging ΛCDM model if not doing better.

**Data Availability Statement** This manuscript has no associated data. [Author's comment: Data sharing not applicable to this article as no datasets were generated or analysed during the current study.]

**Code Availability Statement** This manuscript has no associated code/software. [Author's comment: Code/Software sharing not applicable to this article as no code/software was generated or analysed during the current study.]

## References

1. A.H. Chamseddine, V. Mukhanov, Mimetic dark matter. J. High Energy Phys. **1311**(135) (2013). arXiv:1308.5410 [astro-ph.CO]
2. A. Einstein, The foundation of the general theory of relativity. Annalen de Physik **354**(7), 769 (1916)
3. A. Golovnev, On the recently proposed mimetic dark matter. Phys. Lett. B **728**(135) (2014). arXiv:1310.2790 [gr-qc]
4. L. Sebastiani, S. Vagnozzi, R. Myrzakulov, Mimetic gravity: a review of recent developments and applications to cosmology and astrophysics. Adv. High Energy Phys. 3156915 (2017). arXiv:1612.08661 [gr-qc]
5. A.O. Barvinsky, Dark matter as a ghost free conformal extension of Einstein theory. JCAP **01**, 014 (2014). arXiv:1311.3111 [hep-th]
6. M. Chaichian, J. Kluson, M. Oksanen, A. Tureanu, Mimetic dark matter, ghost instability and a mimetic tensor–vector–scalar gravity. JHEP **12**, 102 (2014). arXiv:1404.4008 [hep-th]
7. J.-W. Lee, Brief history of ultra-light scalar dark matter models. EPJ Web. Conf. **168**, 06005 (2018). arXiv:1704.05057 [astro-ph.CO]
8. L. Hui, Wave dark matter. Annu. Rev. Astron. Astrophys. **59** (2021). arXiv:2101.11735 [astro-ph.CO]
9. P.H. Chavanis, Cosmology with a stiff matter era. Phys. Rev. D **92**, 103004 (2015). arXiv:1412.0743 [gr-qc]
10. S.D. Odintsov, V.K. Oikonomou, Early-time cosmology with stiff era from modified gravity. Phys. Rev. D **96**, 104059 (2017). arXiv:1711.04571 [gr-qc]
11. B.C. Leslie, COSMOLOGY NOW: Lambda-cdm model, What we know and How we know it (2020). ISBN:979-8654852007
12. R. Oerter, The theory of almost everything: the standard model, the unsung triumph of modern physics. (Kindle ed.) Penguin Group (2006). ISBN:978-0-13-236678-6
13. R.J. Nemiroff, B. Patla, Adventures in Friedmann cosmology: a detailed expansion of the cosmological Friedmann equations. Am. J. Phys. **76**(3), 265–276 (2008). arXiv:astro-ph/0703739
14. V. Faraoni, G. Vachan, Quasi-geodesics in relativistic gravity. Eur. Phys. J. C **81**, 22 (2021). arXiv:2011.05891 [gr-dc]
15. F. Capela, S. Ramazanov, Modified dust and the small scale crisis in CDM. JCAP **04**, 051 (2015). arXiv:1412.2051 [astro-ph.CO]
16. T. Buchert, A. Dominguez, Adhesive gravitational clustering. Astron. Astrophys. **438**, 443–460 (2005). arXiv:astro-ph/0502318
17. J. Dutta et al., Cosmological dynamics of mimetic gravity. JCAP **02**, 041 (2018). arXiv:1711.07290 [gr-qc]
18. A.H. Chamseddine, V. Mukhanov, A. Vikman, Cosmology with mimetic matter. JCAP **06**, 017 (2014). arXiv:1308.5410 [astro-ph.CO]
19. P. Ade et al., Planck 2015 results. XIII. Cosmological parameters. Astron. Astrophys. **594**(13), A13 (2016). arXiv:1502.01589 [astro-ph.CO]
20. J.S. Bullock, Notes on the missing satellites problem (2010). arXiv:1009.4505 [astro-ph.CO]
21. W.J.G. de Blok, The core-cusp problem. Adv. Astron. **2010**, 1–14 (2009). arXiv:0910.3538 [astro-ph.CO]
22. P.J. Steinhardt, *Critical Problems in Physics* (Princeton University Press, Princeton, 1997)
23. H. Velten, R.F. vom Marttens, W. Zimdahl, Aspects of the cosmological coincidence problem. EPJC **74**, 3160 (2014). arXiv:1410.2509 [astro-ph.CO]
24. M. Kunz, S. Nesseris, I. Sawick, Constraints on dark-matter properties from large-scale structure. Phys. Rev. D **94**, 023510 (2016)
25. D.B. Thomas, M. Kopp, C. Skordis, Constraining dark matter properties with Cosmic Microwave Background observations (2016)
26. S. Kunar, L. Xu, Observational constraints on variable equation of state parameters of dark matter and dark energy after Planck. Phys. Lett. B **737**, 244–247 (2014)

27. L. Yang et al., Constraining equation of state of dark matter including weak gravitational lensing. Chin. Phys. Lett. **32**, 059801 (2015)
28. W. Yang, H. Li, Y. Wu, J. Lu, Cosmological implications of the dark matter equation of state. Int. J. Mod. Phys. D **26**(03), 1750013 (2017)
29. A. Vilenkin, Birth of inflationary universes. Phys. Rev. D **27**(12), 2848–2855 (1983)
30. S. Tsujikawa, Introductory review of cosmic inflation (2003). arXiv:hep-ph/0304257